\documentclass[twocolumn]{aastex63}

\hypersetup{citecolor=blue, filecolor=blue, urlcolor=blue}

\usepackage{braket}
\usepackage{amsmath}
\usepackage{bm}
\usepackage{enumitem}
\usepackage{rotating}
\usepackage{lineno}

\newcommand{\hb}{H$\beta\,$}
\newcommand{\hgamma}{H$\gamma\,$}

\newcommand{\heii}{\ion{He}{2}}
\newcommand{\cii}{[\ion{C}{2}]}
\newcommand{\civ}{\ion{C}{4}}
\newcommand{\mgii}{\ion{Mg}{2}}
\newcommand{\siiv}{\ion{Si}{4}}
\newcommand{\siii}{\ion{Si}{2}}
\newcommand{\oiii}{[\ion{O}{3}]}

\revised{\today}
\shorttitle{JWST/NIRCam observations of J0100+2802}
\shortauthors{A.-C. Eilers et al.}

\begin{document}\sloppy\sloppypar\raggedbottom\frenchspacing
  
\title{\textbf{EIGER III. JWST/NIRCam observations of the ultra-luminous high-redshift quasar J0100+2802}}

\author[0000-0003-2895-6218]{Anna-Christina Eilers}\thanks{Pappalardo Fellow}
\affiliation{MIT Kavli Institute for Astrophysics and Space Research, 77 Massachusetts Avenue, Cambridge, 02139, Massachusetts, USA}

\author[0000-0003-3769-9559]{Robert A.\ Simcoe}
\affiliation{MIT Kavli Institute for Astrophysics and Space Research, 77 Massachusetts Avenue, Cambridge, 02139, Massachusetts, USA}

\author[0000-0002-5367-8021]{Minghao Yue}
\affiliation{MIT Kavli Institute for Astrophysics and Space Research, 77 Massachusetts Avenue, Cambridge, 02139, Massachusetts, USA}

\author[0000-0003-0417-385X]{Ruari Mackenzie}
\affiliation{Department of Physics, ETH Z{\"u}rich, Wolfgang-Pauli-Strasse 27, Z{\"u}rich, 8093, Switzerland}

\author[0000-0003-2871-127X]{Jorryt Matthee}
\affiliation{Department of Physics, ETH Z{\"u}rich, Wolfgang-Pauli-Strasse 27, Z{\"u}rich, 8093, Switzerland}

\author[0000-0001-8986-5235]{Dominika {\v D}urov{\v c}{\'i}kov{\'a}}
\affiliation{MIT Kavli Institute for Astrophysics and Space Research, 77 Massachusetts Avenue, Cambridge, 02139, Massachusetts, USA}

\author[0000-0001-9044-1747]{Daichi Kashino}
\affiliation{Institute for Advanced Research, Nagoya University, Nagoya 464-8601, Japan}
\affiliation{Department of Physics, Graduate School of Science, Nagoya University, Nagoya 464-8602, Japan}

\author[0000-0002-3120-7173]{Rongmon Bordoloi}
\affiliation{Department of Physics, North Carolina State University, Raleigh, 27695, North Carolina, USA}

\author[0000-0002-6423-3597]{Simon J.~Lilly}
\affiliation{Department of Physics, ETH Z{\"u}rich, Wolfgang-Pauli-Strasse 27, Z{\"u}rich, 8093, Switzerland}

\correspondingauthor{Anna-Christina Eilers}
\email{eilers@mit.edu}

%% TO DO for revision!! 
%% Add this reference \citet{Coatman2019} to discussion: Fig. 6 shows Lbol vs. [OIII]!!! Add Matthew Temple to the acknowledgements

%\linenumbers
\begin{abstract}
We present the first rest-frame optical spectrum of a high-redshift quasar observed with JWST/NIRCam in Wide Field Slitless (WFSS) mode. The observed quasar, J0100+2802, is the most luminous quasar known at $z>6$. We measure the mass of the central supermassive black hole (SMBH) by means of the rest-frame optical \hb\ emission line, and find consistent mass measurements of the quasar's SMBH of $M_\bullet\approx10^{10}\,M_\sun$ when compared to the estimates based on the properties of rest-frame UV emission lines \civ\ and \mgii, which are accessible from ground-based observatories. To this end, we also present a newly reduced rest-frame UV spectrum of the quasar observed with X-Shooter/VLT and FIRE/Magellan for a total of $16.8$~hours. 
We readdress the question whether this ultra-luminous quasar could be effected by strong gravitational lensing making use of the diffraction limited NIRCam images in three different wide band filters (F115W, F200W, F356W), which improves the achieved spatial resolution compared to previous images taken with the \textit{Hubble Space Telescope} by a factor of two. We do not find any evidence for a foreground deflecting galaxy, nor for multiple images of the quasar, and determine the probability for magnification due to strong gravitational lensing with image separations below the diffraction limit of $\Delta\theta\lesssim 0.05\arcsec$ to be $\lesssim 2.2\times 10^{-3}$. 
Our observations therefore confirm that this quasar hosts a ten billion solar mass black hole less than $1$~Gyr after the Big Bang, which is challenging to explain with current black hole formation models. 
\end{abstract}

\keywords{dark ages, early universe --- quasars: emission lines, supermassive black holes --- methods: data analysis --- gravitational lensing: strong gravitational lensing}

\section{Introduction}

Since the discovery of the first luminous quasars nearly six decades ago \citep{Schmidt1965} astronomers have detected and observed more than a million quasars in the universe up to redshifts of $z\gtrsim 7.5$ \citep[e.g.][]{Banados2016, Lyke2020, Wang2021}. Quasars are the most luminous, non-transient objects in the universe, powered by accretion onto a central supermassive black hole (SMBH) of millions to several billions of solar masses in size \citep[e.g.][]{Mazzucchelli2017, Yang2021, WuShen2022}. The most luminous, high-redshift quasar ($z>6$) known to date was discovered by \citet{Wu2015}, i.e.\ J0100+2802 at a redshift of $z=6.3270\pm0.0005$ \citep{Wang2019_J0100}. Its extreme luminosity of $L_{\rm bol}\sim10^{48}\rm\, erg\,s^{-1}$ implies that the quasar hosts a highly accreting SMBH in its center with a mass of $M_\bullet\sim10^{10}\,M_\sun$ \citep{Wu2015}. 

Thus, J0100+2802 harbors the most massive SMBH known at $z\gtrsim 6$ -- which is among the $0.2\%$ of the most massive SMBHs at \textit{all} redshifts \citep{WuShen2022} -- at a time when the universe is only $\sim800$~Myr old. Its existence poses significant challenges to current models aiming to explain the rapid growth of SMBHs in the early universe \citep[e.g.][]{Banados2018}. Adding to the challenge, the quasar's rest-frame ultraviolet (UV) spectrum exhibits a very small proximity zone in the vicinity of the quasar, which indicates a short UV luminous quasar lifetime of $t_{\rm Q}\sim 10^5$~yr, during which accretion onto the black hole is expected to occur \citep[e.g.][]{Eilers2017a, Davies2020, Morey2021}. 

To explain the anomaly of J0100+2802's SMBH it has been suspected that this ultra-luminous quasar could possibly be strongly gravitationally lensed. Arguments in favor of this hypothesis were largely based on the clumpy morphology of observations with the Atacama Large Millimetre Array \citep[ALMA;][]{Fujimoto2020}. However, high-resolution images taken with the Hubble Space Telescope (HST) did not reveal any evidence for multiple images due to strong gravitational lensing \citep[][Yue et al. in prep.]{Fujimoto2020}. Furthermore, detailed analyses of the flux transmission in the quasar's proximity zone \citep{Davies2020}, the expected ratios of X-ray luminosity to rest-frame UV and IR luminosities \citep{Connor2021}, as well as the implications for the  bright-end slope of the quasar luminosity function \citep{PacucciLoeb2020} indicate that a significant magnification of the quasar's luminosity is unlikely. 

The quasar's extremely massive SMBH has also called into question the accuracy of all black hole mass estimates of high-redshift quasars, which are obtained using scaling relations between the width of broad emission lines observed in single-epoch quasar spectra, the quasars' luminosities and their black hole masses \citep[e.g.][]{VestergaardOsmer2009}. These scaling relations are calibrated based on a small sample of quasars at very low redshifts, i.e.\ $z\lesssim 0.2$, for which black hole masses can be precisely determined by means of reverberation mapping (RM) measurements \citep[e.g.][]{VestergaardPeterson2006}. However, RM measurements require long monitoring campaigns and thus for most quasars -- especially at higher redshifts where time delays are longer due to time dilation and their generally more massive black holes -- this method is unfeasible. Thus, masses of SMBHs are inferred from single-epoch spectra using aforementioned scaling relations, most commonly calibrated to the, at low-redshift easily observable, rest-frame optical \hb\ emission line \citep[e.g.][]{VestergaardPeterson2006, Grier2017}. Since the \hb\ line is very challenging to observe for quasars at $z>4$ with ground-based observatories, additional uncertainty and possibly biases in the black hole mass estimates for high-redshift quasars could be introduced by re-scaling the \hb\ emission line properties to properties of rest-frame UV emission lines, such as \civ\ and \mgii\ \citep[e.g.][]{VestergaardOsmer2009, Coatman2017}. These uncertainties have motivated recent efforts to avoid such scaling relations altogether and derive black hole mass estimates and other quasar properties from the spectra themselves by means of data-driven modeling of the quasars' spectral features \citep[e.g.][]{Eilers2022b}. 

In this Letter we present the first spectrum of the ultra-luminous quasar J0100+2802 at rest-frame optical wavelengths observed with the NIRCam grism on JWST, as well as a newly reduced high signal-to-noise spectrum at rest-frame UV wavelengths taken with ground-based facilities (\S~\ref{sec:data}). This spectrum covers the wavelength region between $0.6\mu{\rm m}\leq\lambda\leq 4\mu$m (with a gap between $2.3\mu{\rm m}\leq\lambda\leq 3.1\mu$m), which enables us to obtain measurements of the quasar's black hole mass based on the rest-frame UV \civ\ and \mgii\ broad emission lines, as well as the rest-frame optical \hb\ emission (\S~\ref{sec:BH}). Furthermore, we will show NIRCam Imaging observations that set constraints on the possibility of strong gravitational lensing of the quasar (\S~\ref{sec:lensing}), before summarizing our results (\S~\ref{sec:summary}). Throughout this work, we adopt a flat $\Lambda$CDM cosmology with $\Omega_m=0.31$ and $H_0=67.7~\mathrm{km~s^{-1}~Mpc^{-1}}$ \citep{Planck2018}.

\section{Data}\label{sec:data}

\subsection{Ground-based Spectroscopy} \label{sec:ground-based_data}

The ground-based optical and NIR spectroscopic data for J0100+2802 (RA: $01^{\rm h}00^{\rm m}13^{\rm s}.020$; DEC: $+28^{\circ}02{\arcmin}25{\arcsec}.840$) have been obtained using both the X-Shooter spectrograph \citep{Vernet2011} on the Very Large Telescope (VLT) as well as the Folded-port InfraRed Echellette instrument \citep[FIRE;][]{Simcoe2013} on the Magellan Telescope in the years of 2015 and 2016. The target was observed for a total of $16.8$ hours, of which $11$ hours were observed with VLT/X-Shooter (program ID: 096.A-0095; PI: Pettini) and $5.8$ hours observed with Magellan/FIRE (PI: Simcoe).   

The ground-based optical and NIR data are reduced consistently with the open-source python-based spectroscopic data reduction pipeline \texttt{PypeIt}\footnote{\url{https://github.com/pypeit/PypeIt}} version 1.7.1 \citep{Prochaska2020}. We derive the wavelength solution from the night sky OH lines, in order to have on-sky wavelength calibrations and to reduce the overheads of our observations. The sky subtraction is performed on the 2D images by differencing exposures dithered along the slit in the standard A-B mode and fitting a $b$-spline to further eliminate sky line residuals following \citet{Bochanski2009}. We then perform an optimal extraction \citep{Horne1986} to obtain the 1D spectra. The spectra are flux calibrated using sensitivity functions of standard stars observed during the same observing run. For a few runs no standard stars were observed and hence a sensitivity function is created from an A0 star with known magnitude. 

We co-add the flux-calibrated 1D spectra from each night and correct for telluric absorption features by jointly fitting an atmospheric model and a quasar model. The telluric model grids are produced using the Line-By-Line Radiative Transfer Model \citep[LBLRTM;][]{Clough2005}, while we fit a third order polynomial as the quasar model for this very featureless spectrum. The individual telluric-corrected 1D spectra observed in different observing runs or with different telescopes and instruments are then co-added weighted by the average square SNR of the exposure, and stacked on a common wavelength grid. Note that this steps avoids any interpolation of the data in order to prevent correlated noise properties. 

In order to combine the optical and NIR part of the quasar spectrum observed with X-Shooter, the two parts are stitched together by matching the median flux at the intersection between $9900$~{\AA} and $10100$~{\AA}. As a last step we apply an absolute flux calibration to the fully reduced and co-added quasar spectrum by requiring their integrated flux within the UKIRT-J filter to match the observed $J$-band magnitude of the quasar, i.e. $J_{\rm AB}=17.64\pm0.02$. 

Finally, the spectrum is re-binned to a wavelength grid sampled linearly in velocity space with a chosen step size of $\Delta v \approx 50~\rm km\,s^{-1}$ per pixel. We then divide all native pixels into the respective wavelength bins, and determine the stacked flux in each bin as the mean from all native pixels. Note that the final wavelength grid is the weighted average of the individual wavelengths used for each exposure that fall into a given wavelength bin in the input wavelength grid, and hence not necessarily linearly sampled anymore. 

\begin{figure*}
    \centering
    \includegraphics[width=\textwidth]{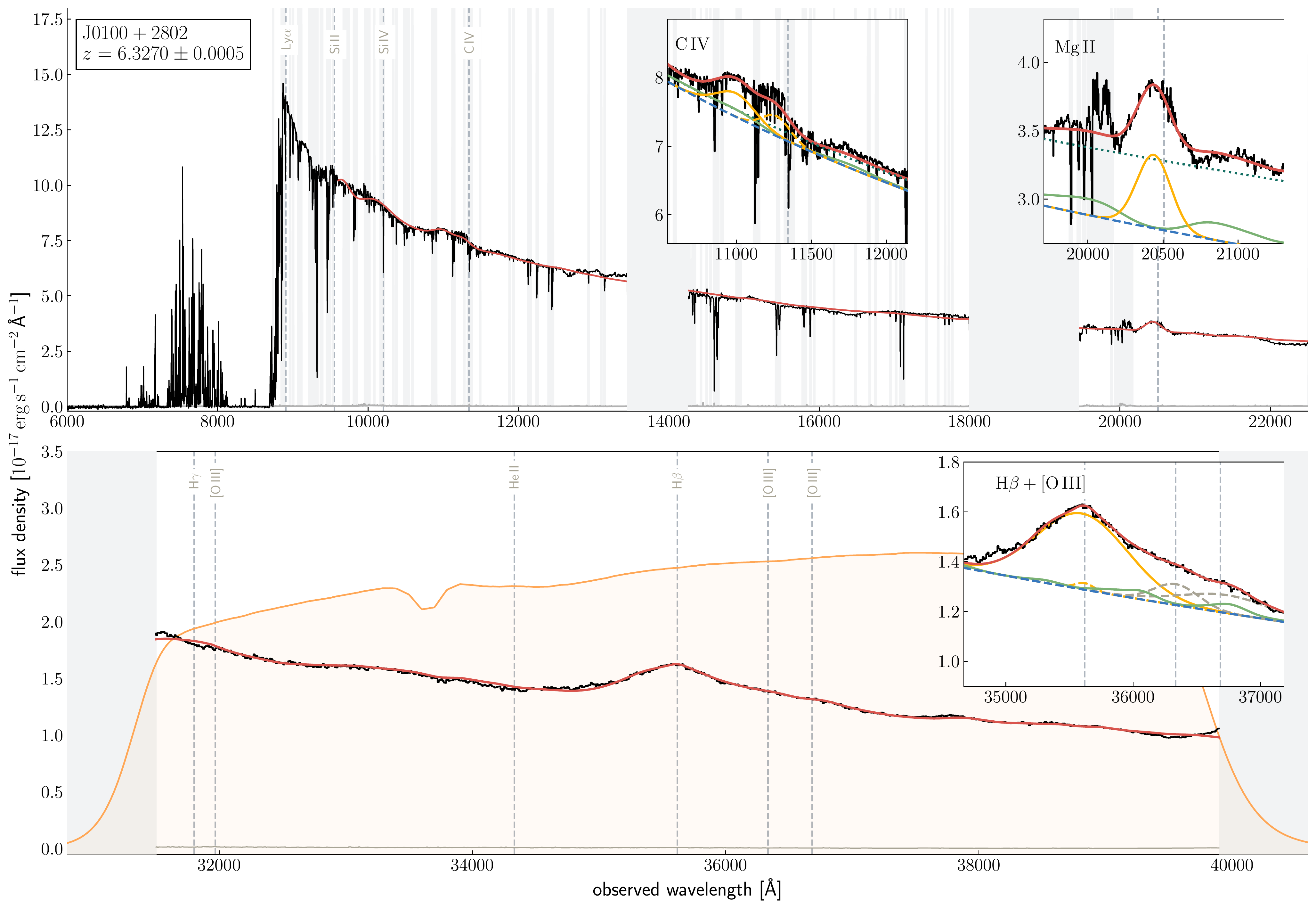}
    \caption{Quasar spectrum of J0100+2802 observed with ground-based facilities (\textit{top}) and NIRCam WFSS on JWST (\textit{bottom}) with the filter F356W shown as the orange shaded region. Grey shaded areas show masked wavelengths affected by foreground absorption systems, regions of significant telluric absorption, regions with strong sky line residuals, as well as regions at the very edge of the filter transmission. The red curve shows the best fit to the quasar spectrum, while the colored curves in the insets show the individual components of the spectral fit (power-law continuum: blue dashed; Balmer continuum: dark green dotted; iron-template: green; Gaussian line components: yellow, grey or purple). }
    \label{fig:spectrum}
\end{figure*}

The resulting ground-based spectrum shown in the top panel of Fig.~\ref{fig:spectrum} has a ${\rm SNR}\approx 205$ per $\Delta v \approx 50~\rm km\,s^{-1}$ pixel measured at a rest-frame wavelength of $\lambda_{\rm rest}=1280\pm20$~{\AA}. 

\subsection{Observations with JWST/NIRCam WFSS}\label{sec:space-based_data}

The quasar field of J0100+2802 was observed with JWST/NIRCam in Imaging and WFSS mode on August 22, 2022, as part of the \textit{Emission-line galaxies and Intergalactic Gas in the Epoch of Reionization} (EIGER) GTO program (program ID: 1243, PI: S.\ Lilly). The quasar field is observed in a mosaic containing four individual visits such that the final field of view spans approximately $3\arcmin\times6\arcmin$ corresponding to $7.6\,{\rm cMpc}\times15.3\,{\rm cMpc}$ at $z\approx 6$. The central $40\arcsec\times40\arcsec$ area around the quasar is part of every visit and thus this region has the deepest observations with a total exposure time of $8760\,$s per visit, i.e. approximately $9.7$~hours in total on source. 
The NIRCam WFSS observations of the J0100+2802 field use the grism ``R'' in the F356W filter. Simultaneously, we obtain direct imaging observations in two short wavelengths filters, i.e. F115W and F200W (see \S~\ref{sec:imaging}). The observational setup as well as the data reduction process are described in detail \citet{Kashino2022} and \citet{Matthee2022}. We will summarize the main steps briefly below. 

The data is reduced using the jwst pipeline\footnote{\url{https://jwst-pipeline.readthedocs.io/en/latest/}} version 1.8.0 provided by STScI. We first run the \texttt{Detector1Pipeline} step to obtain the uncalibrated NIRCam exposures (\texttt{*\_rate.fits}). 
We then apply two additional reduction steps from the \texttt{Spec2Pipeline} to transfer the pixel coordinates to astronomical coordinates (\texttt{AssignWcsStep}) and afterwards apply the flat fielding to the exposures (\texttt{FlatFieldStep}). We then extract the 2D spectrum of the quasar from each of the flat-fielded exposures using customized scripts, which are making use of the \texttt{grismconf}\footnote{\url{https://github.com/npirzkal/GRISMCONF}} module using the latest (V4) trace models. The spectra are normalized by the filter's sensitivity curve and rectified to a common observed wavelength grid, ranging from $3.0\mu{\rm m}\leq\lambda_{\rm obs}\leq4.2\mu$m in bins of $\Delta \lambda=9.75$~{\AA}. 
We correct the trace of the quasar in each exposure for small curvatures and then perform an optimal extraction of the quasar spectrum from each 2D exposure. This results in $96$ individual 1D spectra, which are co-added using the average square SNR of each exposure as weights to a combined 1D spectrum. We estimate the uncertainty on each pixel flux via bootstrapping: to this end we re-sample the $96$ individual spectra with replacement to create $1000$ co-added 1D spectra from which we take the $16$th and $84$th percentile at each pixel as the uncertainty. The final spectrum is shown in the bottom panel of Fig.~\ref{fig:spectrum} and has a ${\rm SNR}\approx 235$ per $9.75 \rm{\AA}$ pixel ($\Delta v\sim80~\rm km\,s^{-1}$) measured at a rest-frame wavelength of $\lambda_{\rm rest}=5100\pm20$~{\AA} . 

The spectrum of the quasar from both the ground-based observatories as well as JWST/NIRCam WFSS will be made publicly available upon publication. 

\subsection{Imaging Data from JWST/NIRCam}\label{sec:imaging}

The NIRCam imaging data was reduced using the \texttt{jwst} pipeline version 1.8.2. In order to avoid saturation in the central pixels of bright sources in the quasar field we run the \texttt{Detector1Pipeline} without suppressing the computations for saturated ramps with only one good, unsaturated sample in the ramp-fitting step. We then run the \texttt{Image2Pipeline} to obtain calibrated images (\texttt{*\_cal.fits}), and apply the $1/f$ noise reduction, remove any snowballs and other artifacts from the data, and subtract the sky and any wisp artifacts. We use \texttt{tweakwcs}\footnote{\url{https://github.com/spacetelescope/tweakwcs}} to align all exposures relative to each other, and then stack all exposure with a global alignment calibrated to Gaia stars. Any offsets are applied to the calibrated images directly, before all exposures are ultimately combined to a final mosaic by the \texttt{Image3Pipeline}. The exposure time of the images in the two short-wavelengths filters F115W and F200W add up to $4380\,$s per visit, i.e.\ approximately $4.9$~hours in total, while the direct images with the F356W filter sum up to $1578\,$s per visit, i.e.\ $1.8$~hours total. Individual exposures are stacked weighted by their inverse variance. 

The final images (for one of the four visits) in the three broad band NIRCam filters F115W, F200W and F356W of a $5\arcsec\times5\arcsec$ region around the quasar are shown in the left column of Fig.~\ref{fig:psf}. Note that we do not take the quasar images from the full mosaic for modeling the PSF (see \S~\ref{sec:lensing}), but only the images from one visit (i.e. visit 1), since it results in a better PSF subtraction. 

Since the noise properties in the short wavelength filters that are reported by the \texttt{jwst} pipeline seem to overestimate the pixel-to-pixel variance in the science image, we re-scale the inverse variance in the science images by a factor of $2.0$ in F115W and $1.8$ in F200W. We do not apply any re-scaling in the long wavelength filter F356W. 

\section{Mass estimates of the quasar's supermassive black hole}\label{sec:BH}

Under the assumption that the dynamics in the quasar’s broad line region (BLR) are dominated by the gravitational pull of the black hole and the system is in virial equilibrium, we can estimate the mass of the SMBH using the FWHM of broad emission lines, which gives an estimate for the velocity of the gas clouds in the BLR orbiting the black hole, and the quasar's luminosity, which acts as a proxy for the radius of the BLR, i.e.  
\begin{equation}
    \frac{M_\bullet}{M_\sun} = 10^\alpha \left(\frac{\rm FWHM_{\rm line}}{1000\,\rm km\,s^{-1}}\right)^\beta\left(\frac{\lambda L_{\lambda}}{10^{44}\,\rm erg\,s^{-1}}\right)^\gamma. \label{eq:bh}
\end{equation}
Traditionally, these scaling relations are calibrated based on a relatively small sample of low-redshift quasars using the properties of the \hb\ emission line. However, as \hb\ is not observable with ground-based observatories for quasars at $z>4$, these scaling relations have been re-calibrated using properties of the rest-frame UV broad emission lines, such as \mgii\ and \civ\ instead. The best-fit values for the scaling parameters $\alpha$, $\beta$ and $\gamma$ dependent on the respective emission lines and monochromatic luminosity $L_\lambda$, and are listed in Tab.~\ref{tab:scaling_relations}. 

\begin{deluxetable*}{lccccc}[!t]
\setlength{\tabcolsep}{12pt}
\renewcommand{\arraystretch}{1.2}
\tablecaption{Parameters of the black hole mass scaling relations. \label{tab:scaling_relations}}
\tablehead{\colhead{emission line} & \dcolhead{\alpha} & \dcolhead{\beta}& \dcolhead{\gamma} & \dcolhead{\lambda}& \colhead{reference}}
\startdata
\civ\ (corrected) & 6.71 & 2 & 0.53 & 1350\,{\AA} & \citet{Coatman2017}\\
\mgii & 6.86 & 2 & 0.50 & 3000\,{\AA} & \citet{VestergaardOsmer2009} \\
\hb & 6.91 & 2 & 0.50 & 5100\,{\AA} & \citet{VestergaardPeterson2006} \\
\enddata
\end{deluxetable*}

\subsection{Spectral Fitting}

To estimate the mass of the SMBH in J0100+2802, we fit the ground-based and NIRCam/WFSS quasar spectrum in the wavelength regions between $0.95~\mu$m and $2.25~\mu$m, as well as $3.15~\mu$m and $3.99~\mu$m, respectively. We mask all wavelengths that are affected by foreground metal absorption systems, regions affected by significant telluric absorption (at $\lambda \approx 14,000$~{\AA} and $\lambda\approx 19,000$~{\AA}), regions with strong sky line residuals (at $\lambda \approx 20,000$~{\AA}), as well as regions at the very edge of the filter transmission (at $\lambda < 31,500$~{\AA} and $\lambda> 40,000$~{\AA}). We fit the observed ground- and space-based spectrum separately, with a combination of (A) a power-law continuum, (B) an iron template (provided by \citet{VestergaardWilkes2001} for the rest-frame UV iron emission and by \citet{Park2022} in the rest-frame optical) that is broadened to match the widths of the broad emission lines through a convolution of the template with a corresponding Gaussian kernel, (C) a Balmer continuum below the Balmer edge \citep[which only applies to the ground-based spectrum; see][for details]{DeRosa2014, Schindler2020}, and (D) one or multiple Gaussian components for the broad emission lines. 

We use a single Gaussian component to fit the \siiv\ $\lambda1304$, the unresolved \siii\ $\lambda\lambda1393, 1402$ doublet, \mgii\ $\lambda2799$, \hgamma\ $\lambda4341$, \oiii\ $\lambda4363$, and \heii\ $\lambda4868$ emission lines, while we use two Gaussian components to capture the more complex line profile of the \civ\ $\lambda\lambda 1548, 1550$, and \hb\ $\lambda4861$ emission lines, as well as the \oiii\ $\lambda\lambda4959,5007$ doublet. This results in $19$ and $32$ free parameters of the spectral fit for the ground-based and NIRCam/WFSS spectrum, respectively, which we estimate using the Markov Chain Monte Carlo (MCMC) algorithm \texttt{emcee} \citep{emcee}. We apply flat priors for each parameter and take the median of the posterior probability distribution as the best parameter estimate. The final spectral fits are shown in red in Fig.~\ref{fig:spectrum}. 

\subsection{Black Hole Mass Estimates}

In order to estimate the mass of the quasar's SMBH using scaling relations, we need to estimate the FWHM of the broad emission lines as well as the monochromatic luminosity $L_\lambda$ (see Eqn.~\ref{eq:bh}). In order to derive the FWHM of the emission lines with multiple Gaussian components, we first combine both profiles to form a joint line profile and measure the width of this joint profile at the half maximum. The estimates for the monochromatic luminosities $L_\lambda$ are determined from the continuum flux $f_\lambda$, for which we evaluate the power-law continuum fit at the respective wavelength $\lambda$ (we evaluate $f_{1350}$ and $f_{3000}$ from the power-law fit to the ground-based spectrum, i.e. $f_{\lambda}\propto\lambda^{-1.6}$, and $f_{5100}$ based on the power-law fit to the grism spectrum, i.e. $f_{\lambda}\propto\lambda^{-2.4}$). The uncertainties on both the FWHM and $L_\lambda$ are estimated from the $16$th and $84$th percentile of $10,000$ random draws from the MCMC posterior.

In order to estimate the black hole mass based on the \civ\ emission line we follow the procedure suggested in \citet{Coatman2017}. It is well known that the \civ\ emission line in high-redshift quasars is often blue shifted due to strong winds and outflows in the BLR \citep[e.g.][]{Meyer2019, Schindler2020}, and thus \citet{Coatman2017} report better consistency in their black hole mass estimates once they ``correct'' the FWHM of the \civ\ line by its velocity shift from the systemic redshift of $6.3270\pm0.0005$ which was determined from the sub-mm \cii\ emission using observations from the Atacama Large Millimetre Array (ALMA) \citep{Wang2019_J0100}. However, determining the velocity shift of the \civ\ emission line is challenging for for J0100+2802 since it is affected by significant foreground and telluric absorption. Thus, to reduce the impact of narrow absorption systems or telluric effects on the emission-line profile we define a ``pseudo continuum'' within the wavelength interval $1450-1600$~{\AA} by applying a median filter to the quasar spectrum, as suggested by \citep{Coatman2016}. Pixels within the wavelength range around the \civ\ emission line profile that lie more than $2\sigma$ below the pseudo-continuum are deemed to be affected by absorption and are additionally masked. All measured spectral properties are listed in Tab.~\ref{tab:spectral_properties}. 

These spectral properties allow us to obtain three different estimates of the black hole mass based on the \civ, \mgii\ and \hb\ emission using Eqn.~\ref{eq:bh}.  Reassuringly, we find consistent results between the black hole mass measurements derived from the rest-frame UV as well as the rest-frame optical emission lines. All estimates point to a SMBH mass of J0100+2802 between $9.7\leq\log_{10}(M_\bullet/M_\odot)\leq10.2$ (see Tab.~\ref{tab:spectral_properties}). The \textit{observational} uncertainties on these measurements are obtained using again the $16$th and $84$th percentile of black hole mass estimates from $10,000$ random draws for the spectral fit parameters from their MCMC posteriors. While the statistical errors on these measurements are small, there are large \textit{systematic} uncertainties in the applied scaling relations of approximately $0.4-0.5$~dex \citep[e.g.][]{VestergaardPeterson2006}, and thus all estimates of the SMBHs are consistent within the expected uncertainties. The measurements show that this quasar indeed hosts a ten billion solar mass black hole at $z=6.3270$, which is the most massive SMBH known in the early universe. 

%XXX SJL:   I think the distinction between what is meant by "statistical" and "systematic" is unclear here (I am certainly no expert in this technique).  Clearly, there is the simple uncertainty introduced by imprecise fitting of the spectra, plus any effects due to finite S/N etc.  I would call these "observational uncertainties" rather than "statistical".   Then there is the scatter in the calibrator sample for each estimator, i.e. for any given set of input data, there is an intrinsic uncertainty in the BH mass when that particular calibrator is applied.  I could imagine that introduces a "statistical uncertainty" for that particular calibrator.  But, what do we know about how these correlate (or not) when different estimators are used?   If there is a high degree of correlation, then I would think "systematic" is reasonable - you do not reduce the uncertainty by using multiple estimators.  If there is not, then I think statistical is better, because you imporve the uncertainty by using multiple estimators.

\begin{figure}
    \centering
    \includegraphics[width=.48\textwidth]{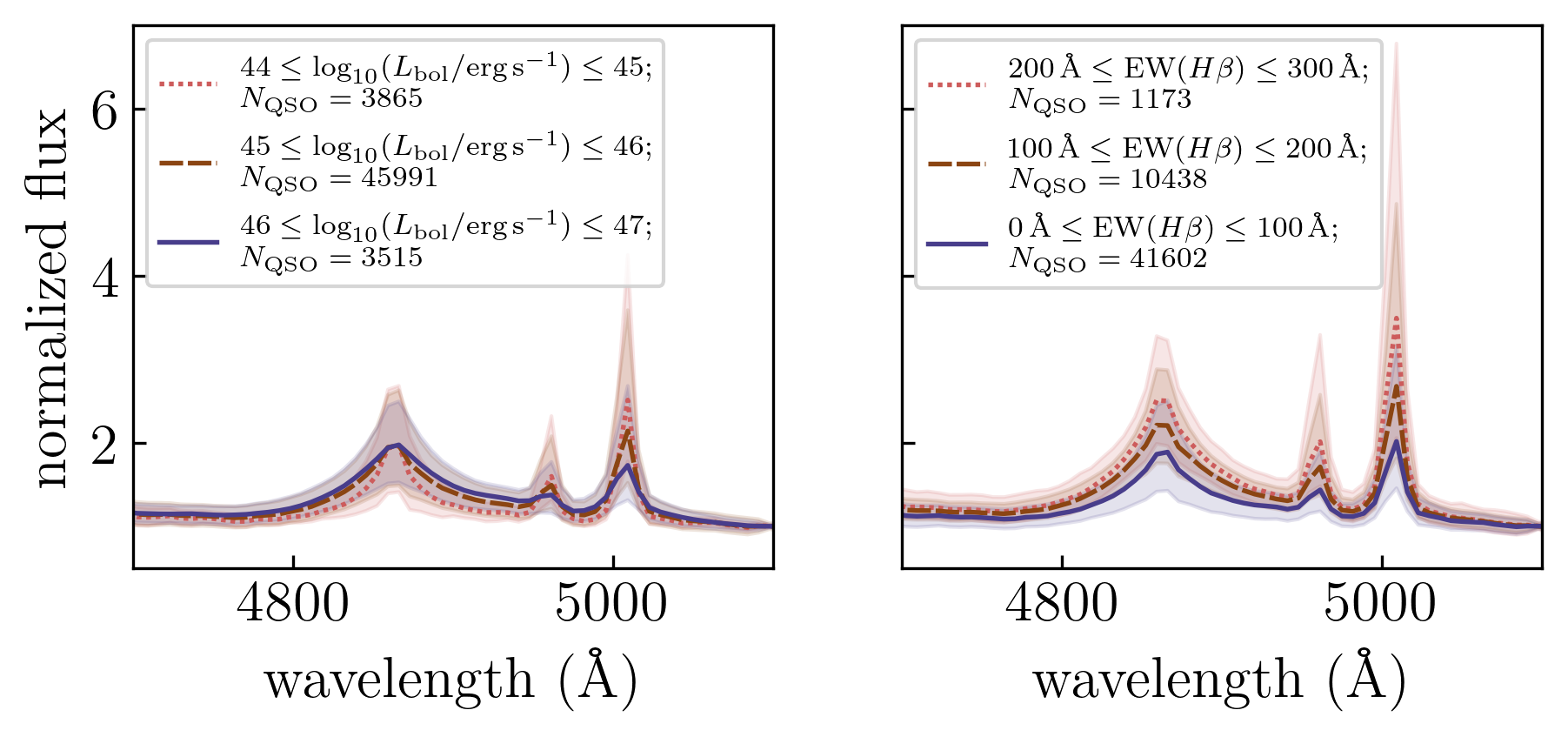}
    \caption{Median rest-frame optical quasar spectra from low-redshift ($0\leq z\leq1$) SDSS quasars, binned by the quasars' bolometric luminosity (\textit{left}) or  EW(\hb) (\textit{right}). All spectra are normalized to unity at $5100$~{\AA}. The spectra show a clear decrease in the strength of the narrow \oiii\ emission lines with increasing $L_{\rm bol}$ and decreasing EW. The quasar J0100+2802 is with a bolometric luminosity of $\log_{10}(L_{\rm bol}/\rm erg\,s^{-1})\approx 48$ more luminous than any of the low-redshift SDSS quasars. However, low-redshift SDSS quasars with a similar EW(\hb) compared to J0100+2802 (i.e. EW(\hb)$=215\pm2$~{\AA}) exhibit clearly narrow \oiii\ emission, which is not present in the spectrum of J0100+2802. }
    \label{fig:sdss}
\end{figure}

\subsection{Equivalent Width of the Quasar's \oiii\ Emission}

We will now take a closer look at the spectral shape of the quasar around the \hb+\oiii\ emission line complex. We estimate an equivalent width (EW) in the observed wavelength frame for the \oiii\ doublet emission of EW(\oiii$)\approx 86\pm1$~{\AA}. %, and an emission line flux of $f_{\rm [O\,III]}\approx 15.7\,\rm erg\,s^{-1}\,cm^{-2}$. 
Interestingly, nearly all of this emission arises from the broad \oiii\ components, since we do not find any narrow \oiii\ emission (i.e. EW(\oiii$_{\rm narrow})<1$~{\AA}) in the spectrum of J0100+2802 (see inset in Fig.~\ref{fig:spectrum}). It is not unusual to observe weaker narrow \oiii\ emission lines with increasing bolometric luminosity \citep[e.g.][]{Baldwin1977} and decreasing equivalent width of the \hb\ emission in low-redshift quasar spectra from the Sloan Digital Sky Survey \citep[SDSS;][]{sdssdr16, Lyke2020}, as shown in the median low-redshift SDSS quasar spectra in Fig.~\ref{fig:sdss}. However, these low-redshift quasars at $0\leq z\leq1$ clearly exhibit narrow \oiii\ emission lines even if the quasars are bright (i.e. $L_{\rm bol}\gtrsim 10^{47}\rm erg\,s^{-1}$), and have a \hb\ equivalent width comparable to J0100+2802 (i.e. EW(\hb)$=215\pm2$~{\AA}). At higher-redshifts, however, this situation changes. \citet{Vietri2018} study the rest-frame UV and optical spectra of 18 hyper-luminous quasars at $2\lesssim z\lesssim 4$ and show that weak \oiii\ emission is common in about $70\%$ of their sample and is strongly correlated with large blueshifts in the \civ\ emission line, as also seen here for J0100+2802. In a similar study using $330$ luminous quasars at $1.5\leq z\leq4.0$ \citet{Coatman2019} confirm the observed anti-correlation between the \oiii\ EW and \civ\ blueshifts, and find that a signifcant fraction of approximately $10\%$ of their quasars exhibit very weak narrow \oiii\ emission (i.e.\ EW$<1$~{\AA}) as seen in J0100+2802, which is $10$ times higher than among the lower-redshift and lower-luminosity SDSS quasars. Thus, the absence of narrow \oiii\ emission in the hyper-luminous quasar J0100+2802 could point towards evolutionary effects in the quasar spectra and might be a sign that the narrow line region (NLR) around the quasar has already been cleared by strong quasar-driven winds on a relatively short timescale \citep{Vietri2018, Coatman2019}. Alternatively, it has been suggested recently that under-luminous \oiii\ emission in quasar spectra might be linked to a low gas content in the NLR of the quasar \citep{Agostino2022}. Upcoming additional rest-frame optical spectra of high-redshift quasars observed with JWST will be required to shed more light on the evolution of the NLR at high redshifts. 

\begin{deluxetable}{lR}[!t]
\renewcommand{\arraystretch}{1.2}
\tablecaption{Spectral properties of the quasar J0100+2802. \label{tab:spectral_properties}}
\tablehead{\colhead{spectral property} & \colhead{measurement}}
\startdata
$\rm FWHM_{\rm C\,IV}~[\rm km\,s^{-1}]$ (corrected) & 3190\pm2070 \\
$\rm FWHM_{\rm Mg\,II}~[\rm km\,s^{-1}]$ & 3890\pm40 \\\vspace{.2 cm} 
$\rm FWHM_{\rm H\beta}~[\rm km\,s^{-1}]$ & 6730\pm170 \\
$\Delta v_{\rm C\,IV}~[\rm km\,s^{-1}]$ & -8850\pm2970 \\
$\Delta v_{\rm Mg\,II}~[\rm km\,s^{-1}]$ & -980\pm20 \\\vspace{.2 cm} 
$\Delta v_{\rm H\beta}~[\rm km\,s^{-1}]$ & -30\pm130 \\
$1350{\rm{\AA}}\,L_{1350\rm{\AA}}~[10^{46}\,\rm erg\,s^{-1}]$ & 40.9\pm0.1 \\
$3000{\rm{\AA}}\,L_{3000\rm{\AA}}~[10^{46}\,\rm erg\,s^{-1}]$ & 25.7\pm0.1 \\\vspace{.2 cm} 
$5100{\rm{\AA}}\,L_{5100\rm{\AA}}~[10^{46}\,\rm erg\,s^{-1}]$ & 20.1\pm0.1 \\
$\log_{10}(M_\bullet / M_\sun)$ (\civ) & 9.6\pm0.9 \\
$\log_{10}(M_\bullet / M_\sun)$ (\mgii) & 9.7\pm0.1 \\
$\log_{10}(M_\bullet / M_\sun)$ (\hb) & 10.2\pm0.1 \\
\enddata
\end{deluxetable}

\section{No Evidence for Strong Gravitational Lensing}\label{sec:lensing} 

\begin{figure*}
    \centering
    \includegraphics[width=.93\textwidth]{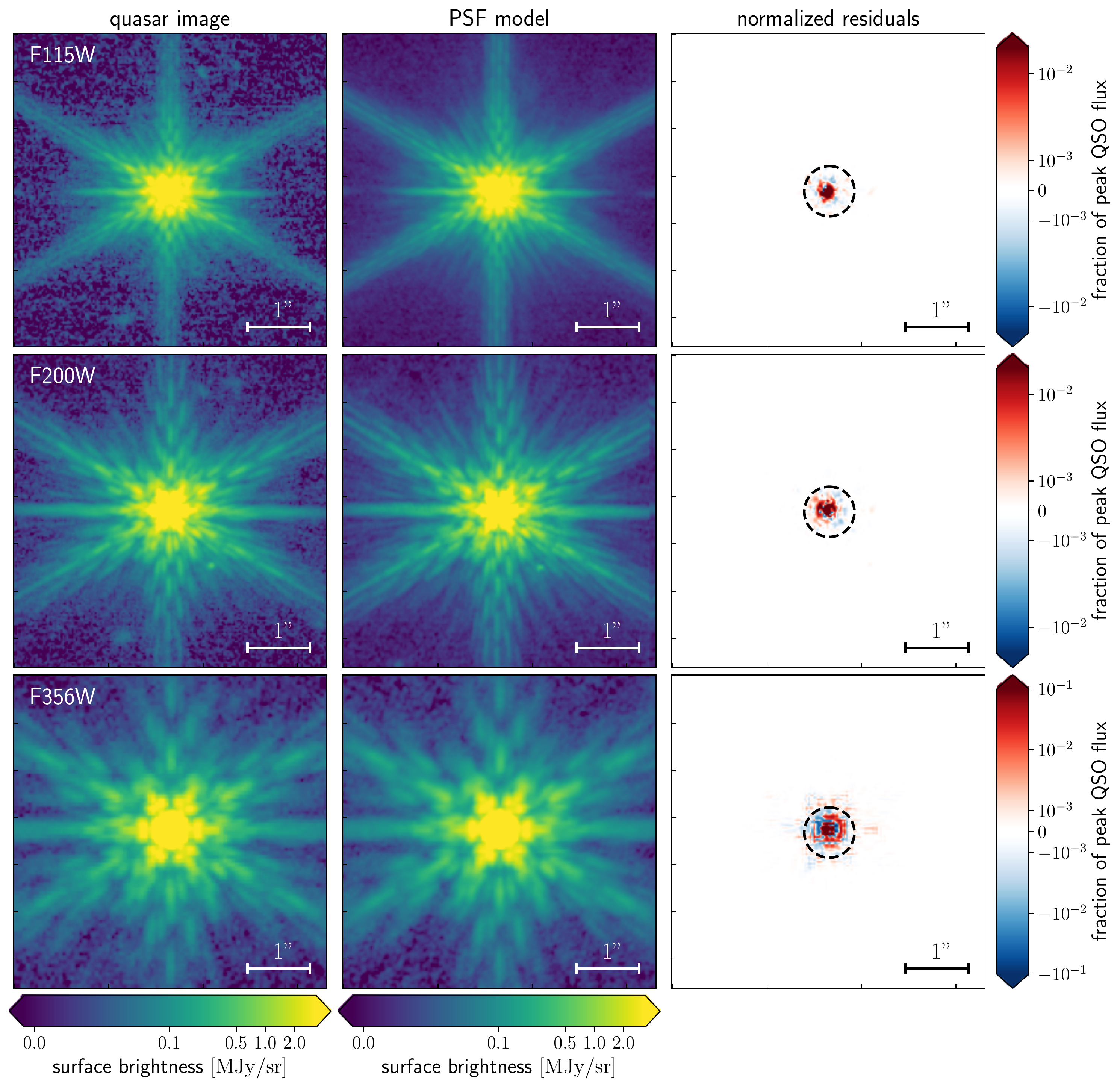}
    \caption{PSF modeling of the quasar images in the NIRCam filters F115W (\textit{top}), F200W (\textit{middle}) and F356W (\textit{bottom}). The left columns show a $5\arcsec\times5\arcsec$ cutouts of the quasars, the middle columns show the quasar model, i.e.\ a single point source convolved with the PSF model, while the right columns show the residuals of the quasar after subtracting the model, normalized by the peak pixel value of the quasar. The black dashed circle indicates the most probable image separation of $\Delta \theta\approx 0.8\arcsec$ for a lensed source at the quasar's redshift (see Fig.~\ref{fig:lensing}). }
    \label{fig:psf}
\end{figure*}

All estimates of the SMBH mass rely on the assumption that the observed luminosity of the quasar is intrinsic to the quasar itself. If the quasar's luminosity was, however, magnified by a factor of $\mu$ due to strong gravitational lensing, the black hole masses would be overestimated by a factor of $\mu^\gamma$ (with $\gamma\approx 0.5$; see Tab.~\ref{tab:scaling_relations}). 

Previous studies searched for multiple images of the quasar or deviations from a single point spread function (PSF) due to strong gravitational lensing effects based on high spatial resolution HST imaging, but have been unsuccessful. The observed quasar images are consistent with a single PSF indicating no evidence for multiple images or arcs with a spatial separation of the diffraction limit of $\theta\approx 0.10\arcsec$, using the F850LP filter at a central wavelength of $\lambda=0.914\,\mu$m on HST's Advanced Camera for Surveys \citep[ACS;][Yue et al. in prep.]{Fujimoto2020}. 

The diameter of JWST's mirror is, at $6.5$~meters, nearly $3$-times larger compared to the $2.4$-meter mirror of HST, which reduces the diffraction limit by a factor of two to $\theta\approx 0.05\arcsec$ in the F115W filter at a central wavelength of $\lambda=1.154\,\mu$m. In the F200W (F356W) filter with a longer central wavelength $\lambda=1.989\,\mu$m ($\lambda=3.568\,\mu$m) the diffraction limit approaches $\theta\approx 0.08\arcsec$ ($\theta\approx 0.14\arcsec$). With the improved spatial resolution of JWST we will now readdress the question, whether we can find any evidence for strong gravitational lensing in this ultra-luminous quasar by detailed modeling of the quasar's PSF.

\subsection{PSF modeling and subtraction}

Making use of the \texttt{photutils} package \citep{photutils}, we build an effective PSF by means of the brightest stars in the quasar field in the respective filter. To this end we first select stars by searching for peaks ($\geq 30~\rm MJy\,sr^{-1}$) in the stacked images, while excluding all peaks that belong to any extended sources rather than stars as well as stars too close ($\leq 2.5\arcsec$) to the edge of our field of view. This results in $5$ ($3$, $4$) stars for the F155W (F200W, F356W) filter from which we build an effective PSF model \citep{AndersonKing2000}. 

Note that since the spectral energy distribution (SED) of the quasar is much redder than the SED of stars, we expect small differences in the shape of the PSFs. However, we nevertheless obtain better results when modeling the PSF when building an effective PSF using stars in the field than when applying \texttt{webbpsf}\footnote{\url{https://webbpsf.readthedocs.io/en/latest/}}, which allows the user to input a quasar SED to construct a PSF model. 

Leveraging the capabilities of \texttt{psfMC}, which performs 2D surface brightness modelling on astronomical images using a MCMC algorithm \citep{Mechtley2014, Marshall2021}, we model the quasar as a single point source and find the best fit PSF model to the quasar images. Note that \texttt{psfMC} requires a noise model for the effective PSF. To this end we assume a SNR$=100$ of the effective PSF at each pixel with a fixed noise floor for all pixels below the $20$th percentile of pixel values.  

The results of the PSF modeling and subtraction are shown in Fig.~\ref{fig:psf} for the three different NIRCam filters F115W, F200W and F356W. The first column shows the quasar images, the second column shows our best fit model, i.e.\ a single point source convolved with the PSF model, while the last column shows the residuals of the quasar after subtracting the PSF model. We do not find any evidence from deviations of a single PSF in any of the three filters. 

In order to further test whether we can find any evidence for a second quasar image, we repeat the modeling procedure and fit the quasar images with two PSFs instead of a single one. The best fit parameters for the central pixels of the two models are less than $1$~pixel (i.e. $0.03\arcsec$) apart, when the magnitudes of both images are approximately equal. Thus we do not find any evidence for two quasar images and conclude that any possible effects due to strong gravitational lensing of the source have to result in image separations of less than a couple of pixels, i.e.\ below the diffraction limit of $\theta\approx0.05\arcsec$.

% Note that these results are independent of our modeling technique, i.e.\ we obtain similar results when modeling the PSF with \texttt{webbpsf}\footnote{\url{https://webbpsf.readthedocs.io/en/latest/}} rather than using the stars in the quasar field, or when fitting the PSF model to the quasar images with \texttt{galfit} \citep{Peng2010} rather than \texttt{psfMC}. 

\subsection{Searching for a foreground deflector galaxy}

We also search for a possible deflector galaxy in the foreground of the quasar, which could cause the quasar to be gravitationally lensed. To this end, we search for any extended emission close to the quasar sightline, since we expect any low-redshift deflector galaxy to be spatially extended given the size distribution of low-redshift galaxies. In \citet{vanderWel2014}, the authors report a median effective radius of $R\approx 1-4$~kpc ($R\approx 3-6$~kpc) for early-type (late-type) galaxies with a stellar mass of $M_\star\sim5\times 10^{10}\,M_\odot$ at redshifts $0<z<3$, which corresponds to spatial extents of $\sim 0.1\arcsec-0.5\arcsec$ ($\sim 0.4\arcsec-0.8\arcsec$). 

% estimate of galaxy size: 
% Lensing cross section is proportional to the mass, most of the mass is in $L^\star$ galaxies, which have a size of $1\arcsec$. 

We do not see any extended sources close to the quasar sightline in the NIRCam filters F115W, F200W and F356W down to the $5\sigma$ maximum sensitivity of 28.7, 29.2, 29.0 magnitude, respectively, in the deepest area of the mosaic around the quasar \citep{Kashino2022}. Furthermore, HST imaging in the filter F606W at an effective wavelength of $\lambda_{\rm eff}=5776.43$~{\AA}, where the quasar's emission is heavily suppressed thanks to the intervening intergalactic medium acting as a natural band pass filter, does not reveal any extended sources around the quasar sightline (Yue et al. in prep.). % down to a limiting magnitude of $XX$ (XX et al. in prep.). 

If a deflecting object would be right along our line-of-sight to the quasar, we would only be able to see the object in the images after subtracting the quasar's PSF, since the quasar would outshine all foreground objects. However, we do not see any significant extended emission around the quasar or nearby, even in the PSF subtracted residual images (right panels of Fig.~\ref{fig:psf}) 

% Could search also in the WFSS data. 
% If the galaxy is still star-forming we should see Paschen emission lines. While the quasar might outshine any foreground emission, we should still see extended emission beyond the quasar PSF. 

\subsection{Probability for strong gravitational lensing}

\begin{figure}[!t]
    \centering
    \includegraphics[width=0.49\textwidth]{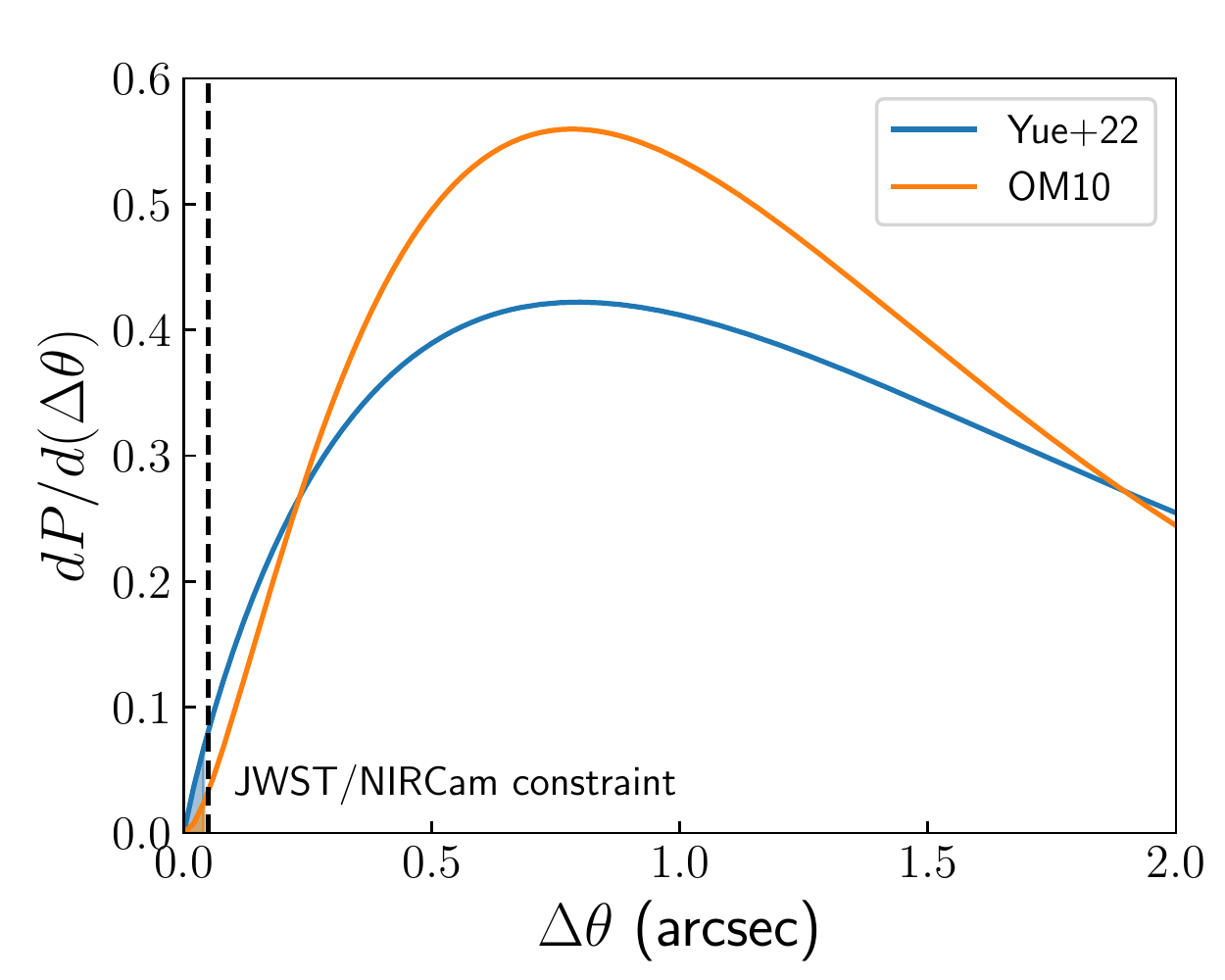}
    \caption{%\textit{Left}: Predicted lensing optical depth $\tau_{\rm m}$ as a function of source redshift $z_{\rm s}$. \textit{Right}: 
    Distribution of the lensing separation $\Delta\theta$ for sources at the quasar's redshift $z_{\rm s} = 6.327$, predicted by the analytical model described in \citet{Yue2022} and \citet[][OM10]{OguriMarshall2010}. The  diffraction limit obtained from the JWST/NIRCam F115W quasar image is shown as the dashed line. }
    \label{fig:lensing}
\end{figure}

We will now estimate the likelihood that the quasar is affected by strong gravitational lensing given the aforementioned constraints on the Einstein radius. We will use two models to describe the population of low-redshift deflector galaxies described in detail in \citet{OguriMarshall2010} and \citet{Yue2022}. Both models use singular isothermal spheres (SIS) to describe the mass profile of the low-redshift deflector galaxies and parameterize the velocity dispersion function (VDF) of the deflectors with a Schechter function \citep{Schechter1976}. However, the VDF applied in \citet{OguriMarshall2010} is based on early-type galaxies derived from SDSS data \citep{Choi2007}, while the VDF in \citet{Yue2022} is based on both late- and early-type galaxies in the local universe \citep{HasanCrocker2019} and assumes a redshift evolution in the parameters of the Schechter function as suggested by \citet{Geng2021}. 
Note that the model by \citet{OguriMarshall2010} predicts a lower number of less massive galaxies compared to \citet{Yue2022}, since the VDF is based on early-type galaxies only (see Fig.~$1$ in \citet{Yue2022} for a comparison between the VDFs). 

Making use of these VDFs of the low-redshift deflector galaxy population, we can built an analytic lensing model of the high-redshift lensed quasar population \citep{Yue2022}. %, which allows us to determine the lensing optical depth of a population of SIS deflectors. The lensing optical depth $\tau_{\rm m}$ describes a prior probability for a source at redshift $z_{\rm s}$ to be multiply imaged \citep{Wyithe2011}, i.e. the fraction of lensed objects among all background sources, and is shown in the left panel of Fig.~\ref{fig:lensing}. We determine this prior lensing probability to be approximately $\tau_{\rm m}\approx 1.8\times 10^{-3}$ for the model by \citet{Yue2022} ($\tau_{\rm m}\approx 2.5\times 10^{-3}$ for the model by \citet{OguriMarshall2010}) for sources at the redshift of J0100+2802, i.e.\ $z=6.327$. 
We show the distribution of image separations of all lensed objects at the quasar's redshift of $z=6.327$ in  Fig.~\ref{fig:lensing}. Both deflector models peak at approximately $\Delta\theta\approx0.8\arcsec$ indicating the most probable image separation. Based on the PSF modeling of the quasar in the JWST/NIRCam images described in the previous section we can constrain the maximum image separation to be below the diffraction limit. Thus, the fraction of lensed objects with image separation $\Delta\theta<0.05\arcsec$ among all lensed objects at $z=6.327$ --- which constitutes the probability that J0100+2802 is gravitationally lensed with an image separation below the diffraction limit --- is $P(\Delta\theta<0.05\arcsec)\approx 2.2\times 10^{-3}$ for the model by \citet{Yue2022}, and $P(\Delta\theta<0.05\arcsec)\approx 6.4\times 10^{-4}$ for the model by \citet{OguriMarshall2010}. %We obtain an estimate for the probability that J0100+2802 is gravitationally lensed with an image separation below the diffraction limit to be approximately $\tau_{\rm m}\times P(\Delta\theta<0.05\arcsec) = 4.1\times 10^{-6}$ for \citet{Yue2022} ($1.6\times 10^{-6}$ for \citet{OguriMarshall2010}). 

%XXX SJL:   IMPORTANT:  I think that this last sentence is not logically correct.    This probabilistic statement surely refers to the probability that a RANDOMLY CHOSEN SIGHTLINE to z = 6.3 is gravitationally lensed with an image separation ....  etc etc.    This particular sightline is certainly not randomly chosen, it has the most luminous quasar known, one for which there are already suggestions that it is lensed.  At the very least, this means that the tau_m factor should be dropped.  I think the most interesting thing that can be done would be to plot a 2-d distribution of image separation vs. magnification for the lens population, and then argue on the basis of that the MAGNIFICATION cannot be larger than X.   My basic point is that you are applying a postiori statistics on a unique system, which is very dangerous.

\section{Summary \& Discussion}\label{sec:summary}

In this letter we present the rest-frame UV as well as the rest-frame optical spectrum at observed wavelengths between $0.6\mu{\rm m}\leq\lambda_{\rm obs}\leq 2.3\mu$m and $3.1\mu{\rm m}\leq\lambda_{\rm obs}\leq 4.0\mu$m, respectively, of the ultra-luminous quasar J0100+2802 at a redshift of $z=6.327$. The ground-based data consist of a total of $16.8$~hours observed with the VLT/X-Shooter spectrograph and Magellan/FIRE, while the rest-frame optical data are observed with JWST/NIRCam in WFSS mode for a total of $9.7$~hours on source. 

We measure the mass of the quasar's SMBH based on the \civ, \mgii\ and \hb\ emission line properties and find estimates that are consistent between the different line estimators within their systematic uncertainties of $0.5$~dex, indicating a SMBH mass of $M_\bullet\sim 10^{10}\,M_\odot$. In the near future, larger samples of rest-frame optical spectra of high-redshift quasars will be able to confirm this long-questioned consistency in black hole mass measurements based on rest-frame UV and rest-frame optical emission line properties for a wider range of black hole masses and quasar properties. 

We revisit the question whether the luminosity of this ultra-luminous quasar is magnified due to strong gravitational lensing with the increased spatial resolution of the JWST/NIRCam images compared to previous HST images, and thus search for evidence for multiple images of the quasars or arcs in three different NIRCam filters, i.e.\ F115W, F200W and F356W. We do not find any evidence from deviations of a single point source at the diffraction limit, nor do we find any evidence for a foreground deflector galaxy. We thus estimate the probability of J0100+2802 being affected by strong gravitational lensing with image separations below the diffraction limit to be $\lesssim2.2\times 10^{-3}$ using two different models for the foreground galaxy population \citep{OguriMarshall2010, Yue2022}. 

Our results confirm that J0100+2802 indeed hosts a SMBH with $M_\bullet\approx 10^{10}\,M_\odot$ at $z=6.327$, when the universe is only about $800$~Myr old, which challenges our current understanding of black hole growth. It requires us to explore models beyond the standard black hole formation paradigm, such as for instance massive initial black hole seeds in excess of stellar remnants \citep[e.g.][]{OhHaiman2002, Begelman2006}, or radiatively inefficient black hole accretion episodes \citep[e.g.][]{BegelmanVolonteri2017, Davies2019b}. JWST has ushered us into a new era for exploring the early universe. Future observations of high-redshift quasars will enable us to study the black hole masses for ensembles of quasars, understand their large-scale environments, and measure their luminosity functions to gain new insights onto the properties and origins of SMBH seeds, and their rapid growth at early cosmic times. \\

\acknowledgements

The authors would like to thank Paul Schechter for insightful discussions about gravitational lensing. Furthermore, we would like to Madeline Marshall for helpful discussions on the PSF modeling, and Gisella De Rosa for interesting discussions on quasar evolution. 

This work is based on observations collected at the European Organisation for Astronomical Research in the Southern Hemisphere under ESO programme 096.A-0095. 

This paper includes data gathered with the 6.5 meter Magellan Telescopes located at Las Campanas Observatory, Chile. 

This work is based in part on observations made with the NASA/ESA/CSA James Webb Space Telescope. The data were obtained from the Mikulski Archive for Space Telescopes at the Space Telescope Science Institute, which is operated by the Association of Universities for Research in Astronomy, Inc., under NASA contract NAS 5-03127 for JWST. These observations are associated with program ID $\#1243$. The specific observations analyzed can be accessed via \dataset[10.17909/xbs2-v060]{https://doi.org/10.17909/xbs2-v060}. 

DK has been supported by JSPS KAKENHI Grant Number JP21K13956.

\software{numpy \citep{numpy}, scipy \citep{scipy}, matplotlib \citep{matplotlib}, astropy \citep{astropy}, PypeIt \citep{Prochaska2020}, photutils \citep{photutils}, psfMC \citep{Mechtley2014}}

\bibliography{literatur_hz}

\end{document}